\documentclass[epj]{svjour}
\usepackage{graphics}
\begin{document}
\title{Well defined transition to gel--like aggregates of attractive athermal particles}
\author{D. A. Head
}                     
%
%
\institute{Department of Applied Physics, The University of Tokyo, Hongo 7-3-1, Bunkyo-ku, Tokyo 113-8656, Japan}
\date{Received: date / Revised version: date}
%
\abstract{
In an attempt to extend the range of model jamming transitions, we simulate systems of athermal particles which attract when slightly overlapping. Following from recent work on purely repulsive systems, dynamics are neglected and relaxation performed {\em via} a potential energy minimisation algorithm. Our central finding is of a transition to a low--density tensile solid which is sharp in the limit of infinite system size. The critical density depends on the range of the attractive regime in the pair--potential. Furthermore, solidity is shown to be related to the coordination number of the packing according to the approximate constraint--counting scheme known as Maxwell counting, although more corrections need to be considered than with the repulsive--only case, as explained. We finish by discussing how the numerical difficulties encountered in this work could be overcome in future studies.
} 
\maketitle

\section{Introduction}
\label{s:intro}

The mechanical properties of athermal aggregates such as granular media ({\em e.g.} sand), wet foams and emulsions are ultimately related to the elastic interactions between their respective constituents (grains, bubbles, droplets) and the morphology of the contact network~\cite{Bolton1990,Mason1995,Head2001}. However, in contrast to chemical bonding in equilibrium systems, where the network morphology can be derived~\cite{Panyukov1996}, persistent elastic contacts form as a result of kinetic energy dissipation leading to arrest, resulting in a nonÐequilibrium state that is often referred to as `jammed'. The mechanics of the arrested state is thus intrinsically linked to its history and loading, potentially allowing for unusual susceptibility properties when different loads are applied~\cite{Cates1998}.

One approach to understanding the nature of this process is to study model systems. Simulations by O'Hern {\em et al.} on spheres with Hertzian or truncated Hookean interactions~\cite{OHern2002,OHern2003} confirmed earlier predictions that the transitional state delineating solid (`jammed') from non--solid systems is {\em isostatic}~\cite{Moukarzel1998}, this term referring to a tenuous solid where severing just one contact would initiate global break up. Macroscopic quantities such as pressure and elastic moduli were shown to scale above this transition with simple exponents that are independent of dimension. The interpretation of the transition remains unclear: a field--theoretic approach suggests a mixed nature~\cite{Henkes2005}, although a possible central role for elastic stability has also been suggested~\cite{Head2005,Wyart2005}.

Greater insight into jamming as a whole could be gained by expanding the range of transitions studied. This is the purpose of this paper: to investigate a different class of jamming transition using similar techniques to previous work. The key difference here is that the particle interactions are {\em attractive} when just touching, allowing for a transition to disordered solid dominated by {\em tensile} contact forces. Indeed, this is precisly what is found: a well--defined transition to a tensile solid at a lower volume fraction than the repulsive--only case. This is not an entirely abstract exercise, as non--Brownian particles will often have short--ranged attractions, such as from van der Waals~\cite{Valverde2004} or surface tension~\cite{Becu2006} effects. Therefore probing the role of jamming in attractive systems will also help to understand the mechanics of a large class of materials.

\begin{figure}
\centerline{\resizebox{0.85\columnwidth}{!}{\includegraphics{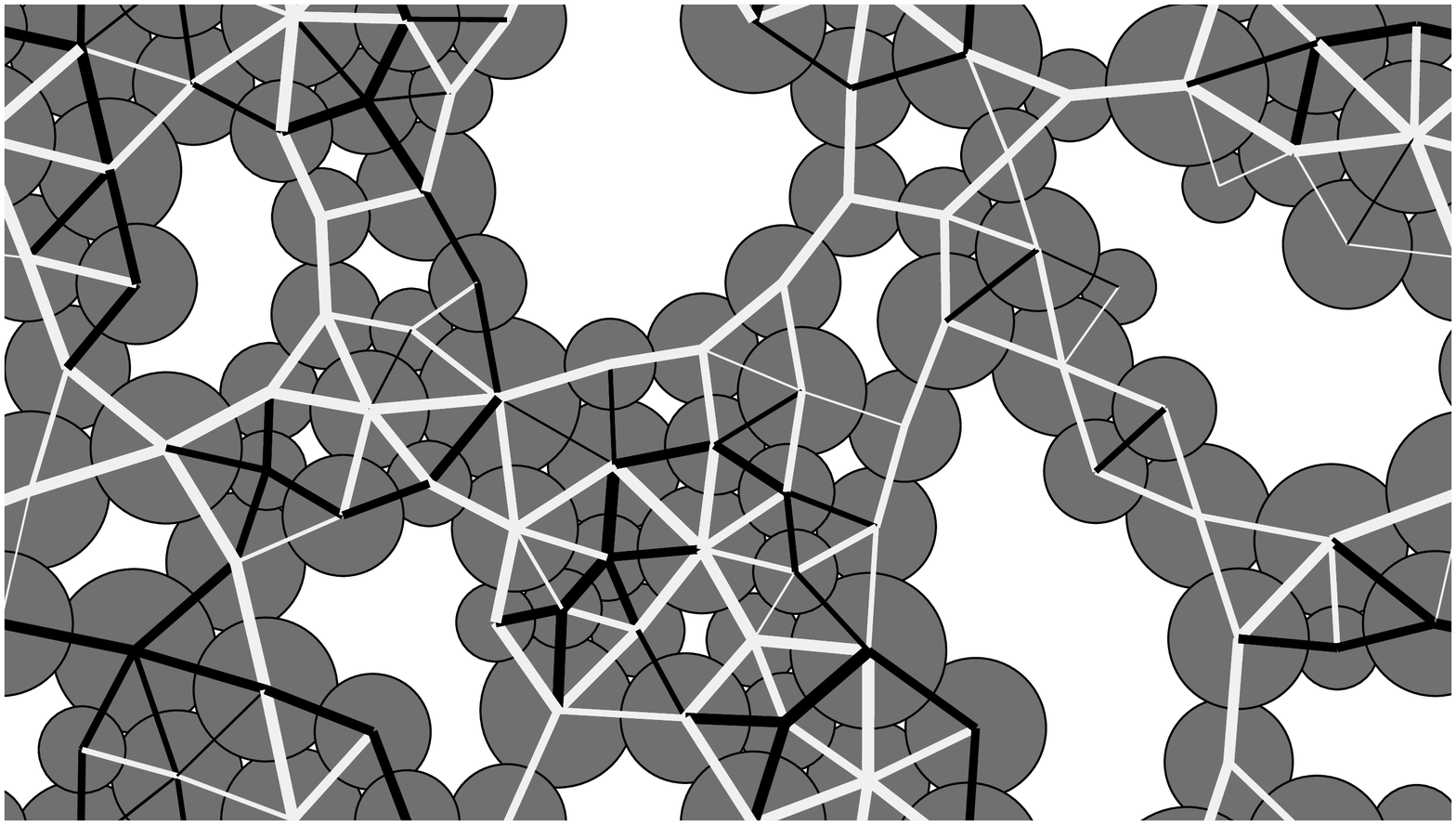}}}
\centerline{\resizebox{0.85\columnwidth}{!}{\includegraphics{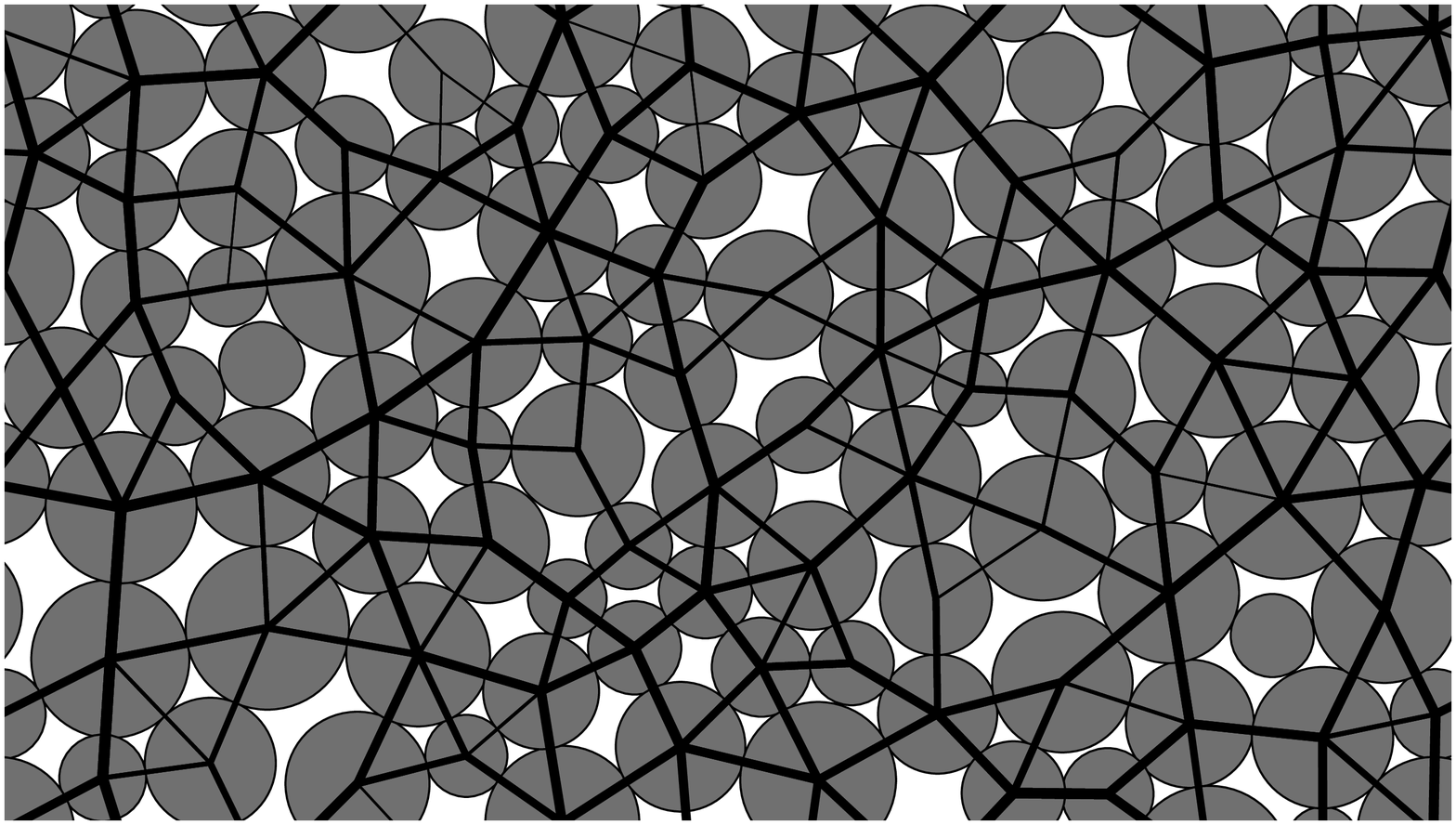}}}
\caption{{\em (Top)}~Example of a relaxed 2--dimensional configuration of athermal particles at mechanical equilibrium, with a radial interaction that is attractive for small overlaps ({\em white lines, thickness proportional to log of force magnitude}) and repulsive for large overlaps ({\em black lines}). For comparison a repulsive--only system at higher density is shown in the lower panel.}
\label{f:sandbox}       
\end{figure}

A visual demonstration of the effects of introducing attractive interactions is shown in Fig.~\ref{f:sandbox}. These snapshots, generated by the freely available demonstration package SandBox~\cite{sandbox:nims}, show polydisperse, 2--dimensional systems of circular particles with the same mixed attractive/re\-pul\-sive potential and initial conditions as used in the rest of this paper, simulated {\em via} a dynamical rule similar to that of Kasahara and Nakanishi~\cite{Kasahara2004}. The message from this figure is that attractive systems tend to cluster into a heterogeneous network with large regions devoid of particles, unlike the much more homogeneous repulsive--only system (also shown). It should therefore not be surprising that solidification occurs at lower densities. Repeated simulations clearly demonstrate the system relaxes to a static state with a negative pressure for a range of densities.

For simplicity, and to facilitate a greater correspondence with previous works, the results presented in the remainder of this paper were generated by an energy minimization algorithm which has no proper mapping to a damped system (indeed, it has no time scale). It is therefore important to stress that, although we find net--tensile states in the $N\rightarrow\infty$ limit with this algorithm, we as yet have no strong evidence this will remain true with dynamical relaxation rules. (The demonstration package described above is limited to systems of around $N=200$ particles). Determining the thermodynamic limit with dynamical relaxation rules is, however, beyond the scope of this article, and must await future verification.

This paper is arranged as follows. In Sec.~\ref{s:model} the model and simulation method are introduced. The results are detailed in Sec.~\ref{s:results} in three parts: firstly the variation of pressure with volume fraction, which shows the existence of the tensile solid. Secondly the variation with system size around this transition point, to confirm the transition is sharply--defined in the infinite system limit. Finally the variation of various coordination numbers with density and how this relates to constraint counting measures of rigidity. Further discussion, including possible directions for future work, is provided in Sec.~\ref{s:disc}.

%
%
\section{Model}
\label{s:model}

An obvious starting point when modelling attractive particles is to select a suitable interaction potential for the system in question. However, this runs the danger of being non--generic, and in the case of surface tension, the standard potentials in use, such as JKR (Johnson, Kendall and Roberts) and DMT (Derjaguin, Muller and Toporov)~\cite{Horn1987}, are non--conservative and hence inappropriate for the energy minimisation algorithm to be used. Therefore a simple shifted and truncated Hookean potential will be used.

The $d=3$ dimensional system comprises of point particles interacting {\em via} with a purely radial pair potential $U(r)$, with $r$ the scalar particle separation. $U(r)$ is strictly zero for particles separated by distances $r>2R$, with the maximum interaction range $R$ is the same for each particle. In terms of the overlap $\varepsilon=2R-r$, 

\begin{equation}
U(\varepsilon)=\left\{
\begin{array}{c@{\quad:\quad}l}
\frac{1}{2}k(\varepsilon-\varepsilon_{\rm shift})^{2}-U_{\rm co} & \varepsilon>0\\
0 & \varepsilon<0
\end{array}
\right.
\label{e:potential}
\end{equation}

\noindent{}where the energy offset $U_{\rm co}=\frac{1}{2}k\varepsilon_{\rm shift}^{2}$ to ensure continuity at $\varepsilon=0$. Thus, there is an attractive regime for small overlaps $0\leq\varepsilon<\varepsilon_{\rm shift}$ and a repulsive regime for larger overlaps $\varepsilon>\varepsilon_{\rm shift}$. A repulsive--only truncated Hookean interaction is recovered by setting $\varepsilon_{\rm shift}=0$.

The system is prepared according to an easily reproducible $T=\infty$ quench protocol: $N$ particles are added at random to an $L\times L\times L$ cubic cell, where $L$ is chosen to give the required volume fraction $\phi=Nv_{\rm p}/L^{3}$ with $v_{\rm p}$ the volume of a single particle (note that this `bare' volume fraction double--counts overlaps). For the data presented below, $\phi$ is calculated on the basis that the true radius of the particle is the minimum of $U(r)$, but no changes in any trends were observed when using the maximum extent~$R$.

The total potential energy ${\mathcal V}(\{{\bf x}_{i}\})$, with ${\bf x}_{i}$ the coordinates of the particle centres, is then minimised {\em via} a high--dimensional optimisation algorithm, either conjugate gradient or Broyden--Fletcher--Goldfarb--Shanno (BFGS)~\cite{NumOptBook}, which were found to agree except near the rigidity transition, for reasons to be discussed in Sect.~\ref{s:disc}. Although it may be tempting to associate energy minimisation with over--damped dynamics, it should be noted that these algorithms do {\em not} move along lines of steepest descent (which is computationally inefficient). Thus strictly speaking there is no precise physical interpretation of this energy minimisation procedure.

%
%
\section{Results}
\label{s:results}

\subsection{Variation of pressure with volume fraction}

For purely repulsive potentials, the pressure $P$ in the final, relaxed state (which is straightforward to extract from the contact forces~\cite{AllenTildesley}) is identically zero below some critical volume fraction $\phi_{\rm c}\approx0.639$, which is sharply defined in the $N\rightarrow\infty$ limit~\cite{OHern2002}. No particles are touching in this state. For $\phi>\phi_{\rm c}$, however, the system relaxes to form a disordered solid with $P>0$ and finite elastic moduli.

\begin{figure}
\centerline{\resizebox{0.85\columnwidth}{!}{\includegraphics{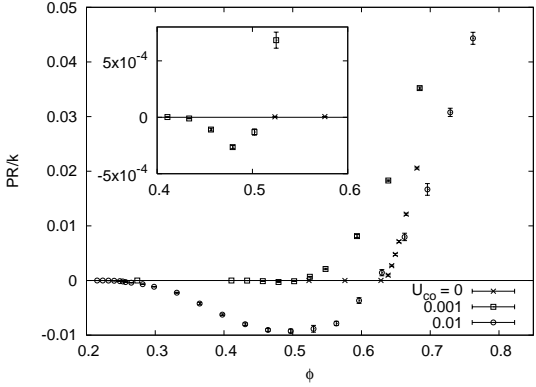}}}
\caption{Plot of normalised pressure $P/(k/R)$ versus the volume fraction $\phi$ for $U_{\rm co}=0$ (repulsive--only), $10^{-3}$ (so $\varepsilon_{\rm shift}/2R\approx 5\%$) and $10^{-2}$ ($\varepsilon_{\rm shift}/2R\approx 15\%$). The system size was $N=1000$ particles. {\em (Inset)}~A magnified region of the data, showing that the $U_{\rm co}=10^{-3}$ data becomes negative. }
\label{f:trend}       
\end{figure}

The introduction of attraction $U_{\rm co}>0$ changes this picture in two qualitative ways, as seen in Fig.~\ref{f:trend}:

(i) $P$ is zero below some critical density $\phi_{\rm c}(U_{\rm co})$ that depends on $U_{\rm co}$, and is {\em negative} immediately above this.

(ii) At some higher density (which also depends on $U_{\rm co}$), $P$ becomes positive.

\noindent{}Numerical convergence was noticeably slower around $\phi_{\rm c}$, which is not surprising as the repulsive--only system is highly correlated here~\cite{Silbert2005}, and the attractive case is expected to be also. Such correlations increase the algorithmic demands in finding a local minimum. No such difficulties arose at the higher density where $P$ crosses over from negative to positive, suggesting there are no unusual susceptibility properties associated with this $P=0$ point.

It is interesting to note that the data for the narrower attractive well, $U_{\rm co}=10^{-3}$, attains net positive pressure $P>0$ at a {\em lower} volume fraction than the strictly repulsive case $U_{\rm co}=0$. This remains true (to a lesser extent) even when the maximum extent $R$ of the particle is used the calculate the particle volumes, rather than the minimum of $U(r)$. Now, given that the $P<0$ regime is weakly gelled (in the sense that $|P|$ is small), it is conceivable that when analysing noisier data such as that gained from experiments, the $P<0$ regime would be entirely missed and `jamming' ascribed to the point when $P$ becomes positive, which is {\em below} random close packing $\phi_{\rm RCP}\approx0.64$. Thus even a narrow attractive regime could significantly contribute to the phenomenon of random {\em loose} packing.

\subsection{Sharpness of the transition}
\label{s:sharpness}

\begin{figure}
\centerline{\resizebox{0.85\columnwidth}{!}{\includegraphics{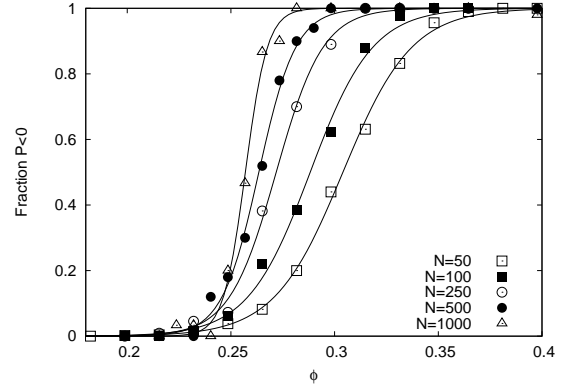}}}
\caption{Fraction of `jammed' systems ({\em i.e.} with $P\neq0$) versus $\phi$ for $U_{\rm co}=10^{-2}$ and the system sizes shown. The solid lines are fits to $\frac{1}{2}\{1+\tanh[(\phi-\phi_{\rm mid})/\sigma]\}$.}
\label{f:frac}
\end{figure}

For the term `transition' to have any real meaning, it must be sharply defined in the
$N\rightarrow\infty$ limit, {\em i.e.} there exists a $\phi_{\rm c}$ such that $P=0$ for all $\phi<\phi_{\rm c}$ and $P\neq0$ for $\phi>\phi_{\rm c}$. For finite~$N$, however, this transition becomes `blurred' and the probability $p_{[P\neq0]}(\phi)$ that a system becomes jammed (here identified by $P\neq0$), varies smoothly from 0 to 1 around $\phi_{\rm mid}\neq\phi_{\rm c}$. Here we focus on the $U_{\rm co}=10^{-2}$ case, and as shown in Fig.~\ref{f:frac} the expected trend of a sharpening transition as $N$ increases is indeed observed.

By fitting $p_{[P\neq0]}(\phi)$ to the functional form

\begin{equation}
p_{[P\neq0]}(\phi)
=
\frac{1}{2}\{1+\tanh[(\phi-\phi_{\rm mid})/\sigma]\}
\end{equation}

\begin{figure}
\centerline{\resizebox{0.85\columnwidth}{!}{\includegraphics{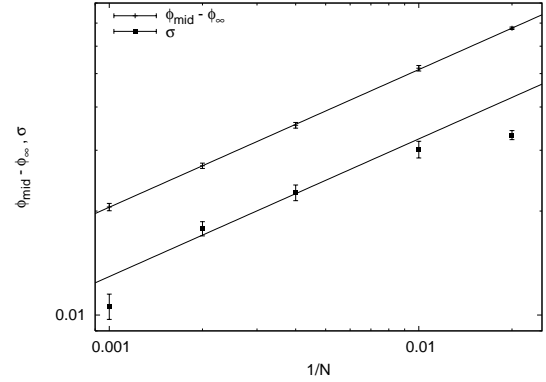}}}
\caption{$\phi_{\rm mid}-\phi_{\infty}$ and $\sigma$ for the fits in Fig.~\ref{f:frac}, plotted against $1/N$. The upper solid line is a fit to (\ref{e:extrapolate}). No convincing fit was found for the noisier $\sigma$ data, including power law, exponential and logarithmic trial forms; the supplied line is a guide to the eye (in fact a fit to $\sigma\propto N^{-\nu}$ with the same $\nu$ as in (\ref{e:extrapolate})).}
\label{f:extrapolate}
\end{figure}

\noindent{}it is possible to extract the width of the transition $\sigma$ and the point when $p_{[P\neq0]}(\phi)=\frac{1}{2}$, $\phi_{\rm mid}$, as a function of system size~$N$, thus allowing the $N\rightarrow\infty$ limit to be extrapolated. This is shown in Fig.~\ref{f:extrapolate}. The data for $\phi_{\rm mid}$ is well fitted by

\begin{equation}
\phi_{\rm mid}
=
\phi_{\infty} + AN^{-\nu}
\label{e:extrapolate}
\end{equation}

\noindent{}with $\phi_{\infty}=\phi_{\rm c}(U_{\rm co}\!\!=\!\!10^{-2})=0.2364\pm0.0017$, $A=0.322\pm0.015$ and $\nu=0.398\pm0.018$. This exponent $\nu$ appears to be consistent with the corresponding repulsive--only value $\approx0.47\pm0.05$~\cite{OHern2003}.

The width $\sigma$ is also a monotonic decreasing function of~$N$, suggesting a sharp transition in the $N\rightarrow\infty$ limit. Although the statistics are poorer for these data and no convincing fit was possible, from the figure it is clearly power law decay or faster. We therefore feel justified in asserting that the transition is sharply defined in the infinite system limit.

\subsection{Coordination number and rigidity}

A cluster can be regarded as solid if its (free) energy increases whenever a non--trivial displacement field is applied to its particles, where `non--trivial' means not rigid body translation or rotation of the cluster as a whole. If all contact forces are initially zero (as typically considered in {\em e.g.} lattice models), the internal energy can only increase if the contact energies increase~\cite{Alexander1998}; for central force interactions as here, this corresponds to a change in the overlaps of contacting particles. The engineering term {\em mechanism}, meaning a displacement mode that does not alter the contact lengths, then coincides with the physics term {\em floppy mode}, meaning a zero-energy mode for which vanishing elastic moduli are expected.

A simple counting argument, known as Maxwell counting, allows the number of mechanisms to be estimated from the network topology. As carefully detailed by Calladine~\cite{Calladine},

\begin{equation}
dN-N_{\rm c}+s=m+\frac{1}{2}d(d+1)
\label{e:maxwell}
\end{equation}

\noindent{}with $d$ the dimension, $N$ the number of particles, $N_{\rm c}$ the number of contacts, $m\geq0$ the number of mechanisms, and $s\geq0$ the number of linearly--independent stressed states. It is normal to express (\ref{e:maxwell}) in terms of the coordination number $z=2N_{\rm c}/N$ and to consider the limit $N\rightarrow\infty$. Then, if we assume only one of $m$ or $s$ can be non--zero at the same time [see (iii) below], $m>0$ only when $z<z_{\rm Maxwell}=2d=6$ for $d=3$, from which we infer the system is floppy. Conversely, $s>0$ when $z\geq z_{\rm Maxwell}$ (so there are `redundant bonds') and we expect a finite restoring force for an arbitrary perturbation.

Thus this counting scheme allows rigidity to be determined from $z$ alone. However, this analysis is only partial:

(i) Transverse contact forces ({\em i.e.} friction), particle asphericity, and non--pair (3 or more) particle potentials change the constraints per contact and degrees of freedom per particle.

(ii) It refers only to the global--spanning rigid cluster; independent subsystems should not be including in the counting. For repulsive--only particles, such subsystems have been identified as individual rattlers~\cite{OHern2003,Donev2005}; for bond--diluted lattice models they are lattice animals that often occupy the bulk of the system~\cite{SahimiBook}.

(iii) Any form of linear dependency between the constraint equations permits $m>0$ and $s>0$ simultaneously, as demonstrated by Calladine~\cite{Calladine}. Such stresses make some mechanisms non--floppy.

(iv) The procedure tells us nothing about the {\em stability} of any solid state~\cite{Head2005,Wyart2005}, and thus may predict solidity for unstable clusters (which will spontaneously rearrange and hence are clearly not solid).

(v) Ordered systems may respond anomalously to specific strains. For instance, a rectangular lattice of unstressed Hookean springs is non--rigid to shears parallel to either of its lattice vectors, but rigid to all others~\cite{BoalBook}.

(vi) Pairs of non--convex particles can make multiple contacts, modifying the counting procedure.

(vii) There is a small error due to the rigid body translation and rotation modes of the cluster as a whole, which must be subtracted from the degrees of freedom. For a free body this correction is $\frac{1}{2}d(d+1)$.

\begin{figure}
\centerline{\resizebox{0.85\columnwidth}{!}{\includegraphics{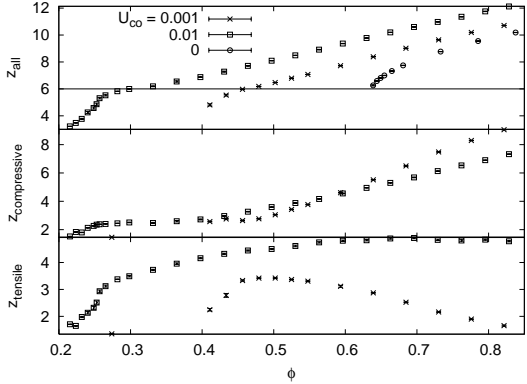}}}
\caption{The coordination number for all contacts $z_{\rm all}$ {\em (upper panel, $z_{\rm Maxwell}=6$ also shown)}, and restricted to compressive and tensile contacts {\em (lower two panels)} for the given $U_{\rm co}$. For this $N=1000$ system, the $U_{\rm co}=10^{-2}$ transition occurs occurs over a range $0.25<\phi<0.27$ (see Fig.~\ref{f:frac}), around where the knee in the data occurs. For clarity, only the $z_{\rm all}=z_{\rm compressive}$ data is given for the repulsive--only case $U_{\rm co}=0$.}
\label{f:z}       
\end{figure}

Plots of $z_{\rm all}=z$ versus $\phi$ are given in Fig.~\ref{f:z}, as well as the same quantity restricted to compressive and tensile bonds, $z_{\rm compressive}$ and $z_{\rm tensile}$ respectively, for various $U_{\rm co}$, including the repulsive--only case $U_{\rm co}=0$. For $U_{\rm co}=0$ the Maxwell approximation is good, {\em i.e.} $P$ becomes non--zero when $z_{\rm all}\approx z_{\rm Maxwell}$. In fact equality can be reached once points (ii) and (vii) above have been corrected for~\cite{Donev2005}.

The introduction of attractions does not appear to drastically alter the picture. For instance, the $U_{\rm co}=10^{-2}$ data, for which the transition occurs over a range $0.25<\phi<0.27$ (see subsection \ref{s:sharpness} above), all three $z$'s exhibit a knee around the transition range, which will presumably become a sharp discontinuity in the thermodynamic limit $N\rightarrow\infty$. The knee corresponds to a sharp increase in $z_{\rm tensile}$, confirming the transition to a tension--dominated state.

Inspection of the raw data suggests $z_{\rm all}\approx5.8$ for jam\-med configurations around $\phi_{\rm c}$ with $U_{\rm co}=10^{-2}$, slightly less than the repulsive--only value $\approx5.86$~\cite{Head2005}. Even taking into account the  correction mentioned in~(vii) (which is $\mathcal{O}(0.01)$ here), this still leaves a significant deviation from $z_{\rm Maxwell}=6$. The obvious explanation is that it is due to a small fraction of rattlers. However, for the attractive case here we must also consider the possibility of independent {\em clusters} of many particles. Indeed, the observation of $z_{\rm all}>0$ for very low volume fractions with $P=0$, which must be due to such clusters, suggests they can make a measurable contribution to $z_{\rm all}$ above $\phi_{\rm c}$ as well.

There is another possible contribution to the gap $z<z_{\rm Maxwell}$ which cannot {\em a priori} be discounted. The existence of self--stressed mechanisms mentioned in (iii) above permits rigid configurations for lower $z$ than expected. For this to be significant, the particles must relax into a configuration in which some constraint equations are linearly dependent. This may appear to be a measure--zero event, but given the stabilising role of such states (see~\cite{Calladine}), they may be attractors of the dynamics and hence arise with finite probabilility. For example, a string of attractive particles pulled taught will naturally form colinear bonds, which are dependent constraints, and such a chain is rigid when under tension (but floppy when unstressed).

%
%
\section{Discussion}
\label{s:disc}

The main finding of this paper has been the demonstration of a well--defined transition to a low--density tensile solid in systems of athermal particles with attractive interactions. This extends the range of model `jamming' transitions that may shed light on the underlying nature of this process. Indeed, if the repulsive--only transition was to be crudely referred to as an athermal {\em glass} transition, the system here could equally crudely be referred to as an athermal {\em gel} transition. Just as in thermal gels (such as colloid--polymer mixtures, see {\em e.g.}~\cite{Dawson2002,Puertas2002}), the non--equilibrium solid forms at low densities, although it is not clear if there is any athermal equivalent of the glass--glass transition.

Only one preparation procedure has been considered here, namely the simple quench from a random, $T=\infty$ state. Whereas repulsive--only particles produce a somewhat homogeneous state, perhaps explaining their robustness to different preparation procedures and the near--ubiquity of `random close packing,' attractive--particle aggregates are far more hetereogeneous, suggesting greater sensitivity to the means of preparation. Indeed, irreversible gelation of emulsions under shear has recently been observed~\cite{Guery2006}, suggesting that jamming could be seen for even lower volume fractions than those observed here if the system is sheared during preparation.

An original intention of this work was to determine the scaling behaviour of various properties, such as pressure or elastic moduli, near the transition point~$\phi_{\rm c}$, as it would be interesting to compare these to the dimension--independent rationals observed in the repulsive case~\cite{OHern2003}. In particular, to see if the visibly less compact nature of the network in Fig.~\ref{f:sandbox} results in fractal spanning clusters at~$\phi_{\rm c}$ and irrational exponents. However, the use of standard minimisation algorithms (conjugate gradient and BFGS~\cite{NumOptBook}) failed due to the non--smoothness of the interaction potential~(\ref{e:potential}), which has a discontinuous first derivative at $\varepsilon=0$ (unlike the strictly repulsive case, which always has a continuous first derivative, and for Hertzian contacts a continuous second derivative also). It proved impossible to make the two algorithms agree near to $\phi_{\rm c}$, and so the $P$ data was discarded here (although $p_{[P\neq0]}(\phi)$ seems to be robust).

Future work could avoid problems with non--smooth potentials by employing a molecular dynamics algorithm, which would also allow the use of standard surface tension interactions~\cite{Horn1987}. It would also be interesting to test the robustness with respect to particle polydispersity and system dimensionality, to see if the exponents remain $d$--independent. If the exponents did differ from the repulsive--only case, one possible interpretation is that the jammed configurations are indeed determined by elastic stability, since stability can be very different under tension than under compression(see {\em e.g.}~\cite{ThompsonHunt}). In this context, it is particularly interesting to note that rational, dimension--in\-de\-pen\-dent exponents are completely typical of elastic instabilities.


%
%
%
\bibliographystyle{epj}
\bibliography{athermal_gel}

\end{document}